# Examining the thermal conductivity of half-Heusler alloy TiNiSn by first-principles calculations

Guangqian Ding, G Y Gao and K L Yao

School of Physics and Wuhan National High Magnetic Field Center, Huazhong University of Science and Technology, Wuhan 430074, China

E-mail: guoying_gao@mail.hust.edu.cn

**Abstract**

The thermoelectric properties of half-Heusler alloy TiNiSn have been studied for decade, however, theoretical report on its thermal conductivity is still little known, because it is difficult to estimate effectively the lattice thermal conductivity. In this work, we use the ShengBTE code developed recently to examine the lattice thermal conductivity of TiNiSn. The calculated lattice thermal conductivity at room temperature is 7.6 W/mK, which is close to the experimental value of 8 W/mK. We also find that the total and lattice thermal conductivities dependent temperature are in good agreement with available experiments, and the total thermal conductivity is dominated by the lattice contribution. The present work is useful for the theoretical prediction of lattice thermal conductivity and the optimization of thermoelectric performance.

(Some figures may appear in colour only in the online journal)

# 1. Introduction

The efficiency of thermoelectric materials is given by the dimensionless figure of merit $ZT=S^2\sigma T/\kappa$, where $S$, $\sigma$, $T$ are the Seebeck coefficient, electrical conductivity, and absolute temperature, respectively. $\kappa$ is the thermal conductivity, which is the sum of contributions from electron ($\kappa_e$) and lattice ($\kappa_l$) parts [1]. High $ZT$ means a high power factor ($PF=S^2\sigma$) and a low $\kappa$. However, due to the restriction in these coefficients, a high $PF$ always comes with a high $\kappa$ in bulk materials such as half-Heusler alloys [2,3]. For a long time, reducing the $\kappa$ has been treating as an independent work in order to look into the possible increase in $ZT$. Much effort mainly focused on reducing the lattice thermal conductivity because conducting works on $\kappa_e$ is in vain for its proportional relationship with $\sigma$. The lattice thermal conductivity is determined by the behavior of phonon transport in the materials. Increasing scattering on phonons would lower the heat flux, and thus result in reducing the thermal conductivity from phonons. Several approaches have been taken to achieve this, including enhancement of impurity scattering by doping, boundary scattering by low dimensions, and carrier scattering by heavy doping. So far, most of the works are blindly conducted in experiments [4,5,6], and theoretical explanation and prediction are still lacking.

TiNiSn-based half-Heusler alloys as high temperature thermoelectric materials have been studied for decade [2-7]. However, its high thermal conductivity always leads to low $ZT$ and thus greatly weaken its promising thermoelectric applications. Surprisingly, the attempts made to reduce its lattice thermal conductivity are found to have less effect on its $PF$ due to its low carrier mobility [8], and thus the possible improvement of $ZT$ is expected. In 2003, Katayama *et al.* [2] reported an upper limit $ZT$=0.3 for single crystal TiNiSn at 750 K with $\kappa$ of 6 W/mK experimentally. In order to improve its *ZT*, they tried to decrease the thermal conductivity by doping TiNiSn with Hf, Zr, Pt, and Si (Hf, Zr on Ti site, Pt on Ni site, and Si on Sn site). They found that thermal conductivity decreases by all the additions except for Si, and a maximum increased *ZT* about 0.4 was obtained for $Ti_{0.9}Hf_{0.1}NiSn$. There are also works mainly conducted on the lattice thermal conductivity. The room temperature $\kappa_l$ of bulk TiNiSn measured by Bhattacharya *et al.* [5,6] was 8W/mK. They also introduced impurity of Sb on Sn site to fabricate $TiNiSn_{1-x}Sb_x$ with different concentrations, unexpectedly, no obvious decrease in the lattice thermal conductivity was found. As for earlier experiment, Kafer *et al*. [4] also intended to reduce $\kappa$ by alloying the ternary half-Heusler alloys. All these works are in experiment and short of combination with theoretical calculation and analysis. So far, available theoretical studies in recent years are only focused on the properties of electronic structure and *PF* [9,10], and research on $\kappa$ is still lacking in theory, because it is difficult to estimate effectively the lattice thermal conductivity. In the present work, we will systematically study the thermal conductivity of basic TiNiSn by the first-principles combined with Boltzmann transport theory, the ShengBTE code [11] developed recently, and detailed comparison with available experiments would be made.

## 2. Computational details

Half-Heusler alloys have a MgAgAs structure with the space group $F\bar{4}3m$. The relaxed lattice constant is 5.89 Å for TiNiSn, which is in good agreement with the experimental value 5.92 Å [4]. Next, the calculation process of $\kappa_e$ and $\kappa_l$ would be performed. As for $\kappa_l$, most of the theoretical models presented by researchers usually begin with Boltzmann transport equation (BTE), which describes the rate of change in the distribution function of phonons under the presence of external fields (temperature gradient, electric field, etc). The earlier typical model we can find is Callaway's model [12]. The scattering processes of phonons in Callaway's model was just classified into *N* processes (normal processes) and *U* processes (umklapp processes, impurity scattering and boundary scattering), and the frequency-dependent relaxation times are used to represent the phonon scattering processes in order to better solve the BTE. This model was later developed by Allen [13]. However, these models including parameters that strongly depend on experimental data, and the difficulties in determining the relaxation times also lack their predictive power in $\kappa_l$. In this work, we use the ShengBTE code [11] based on phonon Boltzmann transport equation (pBTE) and recently presented by Li *et al.* to study the $\kappa_l$ of TiNiSn. The code is based on the second-order (harmonic) and third-order (anharmonic) interatomic force constants (IFCs) combined with a full solution of the pBTE and successfully predict the $\kappa_l$ of materials [14,15]. We use 4×4×4 supercell for calculations of second and third-order IFCs. The second-order IFCs is obtained from the Phonopy package [16], and third-order IFCs is obtained by the thirdorder.py script in ShengBTE [11]. Then, the $\kappa_e$ is calculated by the electronic Boltzmann transport equation (eBTE) within the relaxation time approximation as implemented in the BoltzTraP [17], a package which based on the electronic structure and can give the prediction of *S*, $\sigma$ and $\kappa_e$. It should be mentioned that this package can only give the results with respect to relaxation time $\tau$, and we later estimate the value of $\tau$ by fitting the calculated result with experimental value. Note that all supercells and self-consistent calculations are performed in the first-principles package VASP [18].

## 3. Results and discussion

The phonon dispersion relation calculated from the harmonic force constants along several high-symmetry lines for TiNiSn is shown in Figure 1. Since the primitive cell of TiNiSn contains only three atoms, nine independent vibration modes can be found, of which three are three acoustic modes (two transverse and one longitudinal) and the remaining six are optical ones. In contrast to skutterudites [19], it is found that the acoustic branches of TiNiSn separate from optical ones with a large frequency gap. The maximum frequencies of three acoustic branches ($\omega_{TA}=110 cm^{-1}, \omega_{TA'}=125 cm^{-1}$, and $\omega_{LA}=151 cm^{-1}$) are in good agreement with former results [20]. The frequency of the optical modes approximate to constant within the range of wave vector, leading to low group velocity in these phonons. Thus, just like in most cases, the optical modes will not be interested because their contribution to thermal conductivity is limited. Considering Callaway and Cahill [12,21], whose theoretical mode also ignore the contribution of optical branches. Table 1 shows the estimated acoustic phonon velocities and room temperature $\kappa_l$ of TiNiSn compared with conventional bulk thermoelectric materials [22-24]. The phonon velocities are given by the slope of acoustic phonon dispersion around the $\Gamma$ point. The room temperature $\kappa_l$ is the ShengBTE's result. As shown in Table 1, the average acoustic velocity of TiNiSn is 3824 $m/s$, which is much higher than other bulk materials. According to the description of lattice thermal conductivity ($\kappa_l = \frac{1}{3} C_v v_s l$) borrowed from the dynamical theory of glass molecules [25], $\kappa_l$ is proportional to phonon velocity and mean free path (MFP). High velocity means fast heat transport and a large mean free path guarantee a low phonon scattering rate, both making their contribution to a high lattice thermal conductivity. It is evident that TiNiSn has the largest $\kappa_l$ among these thermoelectric materials.

Combining the second-order and third-order IFCs, the lattice thermal conductivity is calculated in ShengBTE [11]. It is worth noting that the two kinds of IFCs are necessary for the study of heat transport. Under harmonic approximation, the lattice vibration is regarded as harmonic, and we can thus ignore the interaction between phonons. This approximation is just reasonable at low temperature when lattice vibration is much weak. As temperature goes up, severe vibration will increase the coupling of phonons, which would greatly limit the MFP. These two aspects should be considered so that we can get reliable $\kappa_l$ for the whole temperature range. Calculated results for TiNiSn, at temperatures from 100 K to 1000 K, are plotted in Figure 2. The insert figure gives the comparison between our calculated results and Bhattacharya's experimental data (100 K-300 K). We predict the $\kappa_l$ at the room temperature is 7.6 W/mK, which is very close to the value of 8 W/mK obtained by Bhattacharya [5,6]. It is apparent that the $\kappa_l$ is slightly overestimated in comparison with experiment in low temperatures, and better compliance can be found as temperature increases. In a complete bulk crystal, phonon-phonon scattering will act as a major scattering mechanism, especially the three-phonon scattering. High temperature will excite more phonons and keep a high energy of them, which greatly increase the scattering rate and limit the MFP. Thus, throughout the temperature zone, the $\kappa_l$ decreases as temperature goes up. One might wonder whether impurity in real bulk TiNiSn could play a significant role in reducing its thermal conductivity. As can be seen from former experimental measurements [2,5], the result is not satisfactory. The reason why the effect of impurity so small is that the anharmonic phonon-phonon scattering rates are much higher than phonon-point defects scattering rates. Furthermore, point detects only have effect on the high frequency phonons, while this kind of phonons is few and show low velocity in bulk TiNiSn. In order to completely study the thermal conductivity, we still have to research the electronic contribution.

We predict the electronic thermal conductivity of TiNiSn by eBTE within relaxation time approximation [17]. This method has been extensively used to study the electronic thermal conductivity [26]. Under the relaxation time approximation, $S$ is independent of the relaxation time $\tau$, whereas $\sigma$ and the $PF$ ($S^2\sigma$) depend linearly on $\tau$. Furthermore, the $\kappa_e$ is obtained by Wiedemann-Franz equation ($\kappa_e = L\sigma T$, $L$: Lorentz factor), thus, $\kappa_e$ is also relevant to $\tau$. Calculating the relaxation time individually is trivial. Here, the value of $\tau$ is estimated by fitting the electrical resistivity $\rho$ at 300 K to the experimental value of $16.5\times10^{-5}\Omega m$ under the room temperature carrier concentration of $1.3\times10^{19} cm^{-3}$ as measured by Kim *et al.* [3]. This yields the relaxation time $0.9\times10^{-14}$ s. The $\kappa_e$ as well as $\kappa_l$ and the total thermal conductivity ($\kappa_e + \kappa_l$) are plotted in Figure 3. The insert figure gives the experimental thermal conductivity of TiNiSn measured by Kim *et al.* at temperatures from 300 K to 1000 K [3], which perfectly meets our prediction. Comparing $\kappa_e$ with $\kappa_l$ indicates that the thermal conductivity is dominated by lattice contribution. The thermal conductivity decreases with the temperature up to around 650 K and then increases. The increase in thermal conductivity at high temperatures is mainly due to the remarkable increase in electronic contribution. It is well-known that TiNiSn-based half-Heusler alloys always exhibit maximum *ZT* around 700-800 K [2,3,7]. The reason is that the maximum power factor and lowest thermal conductivity occur simultaneously at high temperature.

## 4. Conclusion

In summary, we systematically investigate the thermal conductivity of half-Heusler alloy TiNiSn by first-principles calculations. The $\kappa_e$ and $\kappa_l$ are individually calculated from BTE. The results are good consistent with available experiments. We also show that the total thermal conductivities in TiNiSn are dominated by the lattice contribution. The present work will stimulate theoretical prediction to lattice thermal conductivity and optimization of thermoelectric performance for more half-Heusler thermoelectric materials by using the present first-principles method.

**Acknowledgments**

This work was supported by the National Natural Science Foundation of China under Grant Nos. 11474113, 11004066 and 11274130, and by the Fundamental Research Funds for the Central Universities under Grant No. HUST: 2013QN014.

# Tables

Table 1. Acoustic phonon velocities (two transverse and one longitudinal) and room temperature lattice thermal conductivity of TiNiSn. Corresponding values of three conventional bulk thermoelectric materials of SnSe,[23] $Bi_2Te_3$,[22] and PbTe[24] are also presented.

|  | TA | TA′ | LA | Average | $K$ (300K) |
|---|---|---|---|---|---|
| TiNiSn | 2992 | 3020 | 5462 | 3824 | 7.6 |
| SnSe[23] | 1233 | 1649 | 2355 | 1745 | 0.6 |
| $Bi_2Te_3$[22] | 1590 | 1590 | 2840 | 2006 | 1.2 |
| PbTe[24] | 1610 | 1610 | 3596 | 2272 | 1.5 |

**Figures**

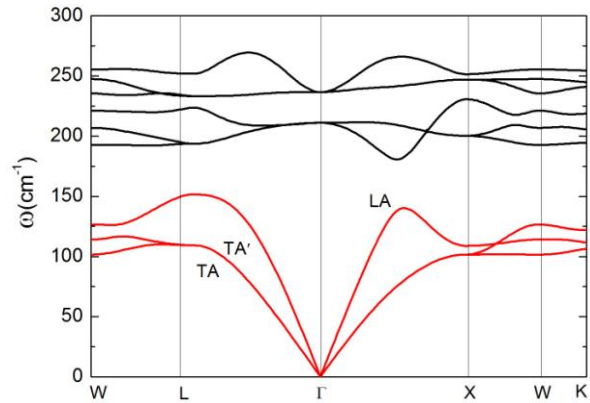

**Figure 1.** Phonon dispersion relation for TiNiSn along several high-symmetry lines calculated using Phonopy package.[16] The acoustic modes are highlighted in red.

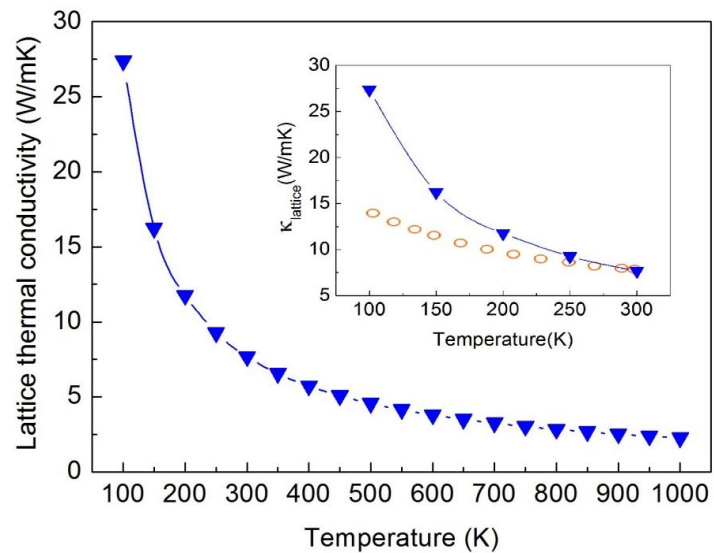

**Figure 2.** Calculated lattice thermal conductivity as a function of temperature for TiNiSn. Insert: the comparison with Bhattacharya's experimental data[5,6] (the line of read hollow cricle) from 100 K to 300 K.

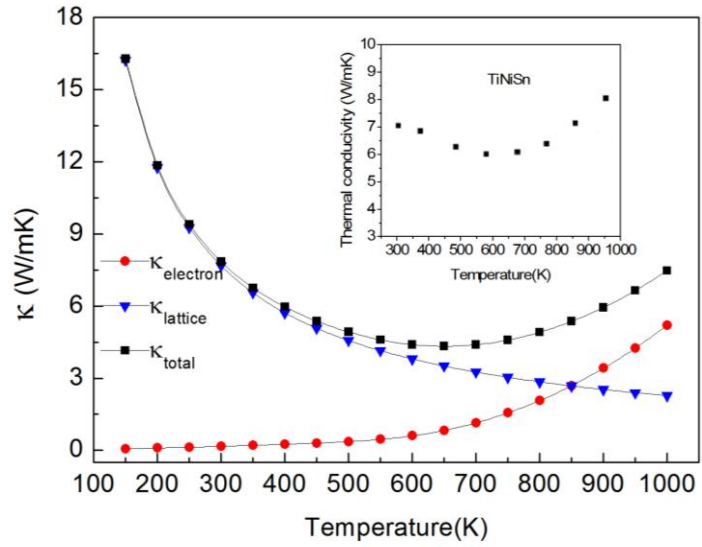

**Figure 3.** The calculated electronic thermal conductivity as well as lattice thermal conductivity and total thermal conductivity for TiNiSn as a function of temperature. Insert: the experimental total thermal conductivity of TiNiSn by Kim *et al.*[2,3] from 300 K to 1000 K.